\documentclass[12pt]{iopart}
\usepackage{iopams,verbatim}

\def\Journal#1#2#3#4{{#4} {\it #1} {\bf #2}, #3 }
\def\p{\partial}
\def\fpd#1#2{\frac{\partial #1}{\partial #2}}

\def\b{\beta}
\def\g{\gamma}

\def\e{\epsilon}
\def\p{\partial}
\def\l{\lambda}

\def\s{\sigma}
\def\t{\theta}
\def\w{\omega}
\def\W{\Omega}
\def\wph{\mu}

\def\di{\textrm{d}}

\begin{document}

\title{A Petrov type I and generically asymmetric rotating dust family}

\author{Lode Wylleman}

\address{Faculty of Engineering Sciences, University of Ghent (UGent), Galglaan 2, 9000 Gent, Belgium}

\ead{lwyllema@cage.ugent.be}

\begin{abstract}
The general line element corresponding to the family of
algebraically general, gravito-electric, expanding, rotating dust
models with one functionally independent zero-order Riemann
invariant is constructed. The isometry group is at most
one-dimensional but generically trivial. It is shown that the
asymmetric~\footnote{without non-trivial isometries} solutions with
constant ratio of energy density and vorticity amplitude provide
first examples of Petrov type I space-times for which the Karlhede
classification requires the computation of the third covariant
derivative of the Riemann tensor.
\end{abstract}


\section{Introduction}


Rotating dust solutions in general relativity may serve to describe
phenomena on a galactic scale. The metric $g_{ab}$ obeys the field
equation
\begin{eqnarray}
R_{ab}-\frac{1}{2}R\,g_{ab}+\Lambda\, g_{ab}=\wph\, u_a\,u_b
\end{eqnarray}
with, as usual, $R_{ab}$ the Ricci tensor, $R$ the Ricci scalar,
$\Lambda$ the cosmological constant, $u^a$ the dust 4-velocity field
and $\wph$ the energy density. For a space-time filled with rotating
dust the fluid flow is non-accelerating and the remaining kinematic
variables are the expansion scalar $\t\equiv u^a{}_{;a}$, shear
tensor $\s_{ab}\equiv u_{(a;b)}-\frac{\t}{3}\,h_{ab}$ and vorticity
vector $\w^a\equiv \frac{1}{2}\e^{abc}u_{[b;c]}\neq 0$ of the fluid,
with $h_{ab}\equiv g_{ab}+u_a u_b$  and
$\e_{abc}\equiv\eta_{abcd}u^d$ the spatial permutation tensor.

Important classes of rotating dust models have been found by
assuming some kind of symmetry, or are algebraically special.
Respective examples are Winicour's classification~\cite{Winicour} of
stationary axisymmetric models satisfying the circularity condition
(see \cite{SKMHH} for examples and further discussion), and the
general rotating dust solution admitting time-like conformally flat
hypersurfaces with zero extrinsic and constant intrinsic curvature
as found by Stephani~\cite{Stephani} and later generalized by Barnes
for non-zero $\Lambda$~\cite{Barnes}, which depends on seven free
functions of one coordinate and which turns out to be of Petrov type
$D$. A final and famous example is the homogeneous Petrov type $D$
G\"{o}del universe~\cite{Godel}, which can be interpreted as a
rotating dust space-time with a negative cosmological constant.

There seems to be, however, a lack of
algebraically general, asymmetric solutions~\footnote{To the best of
my knowledge, no such models have been found so far.}. In a search
for such,
the
class ${\cal A}$ of Petrov
type $I$, \emph{gravito-electric} rotating dust models, i.e.\ for
which the Weyl tensor wrt observers comoving with the dust is purely
electric,
\begin{eqnarray}
H_{ab}\equiv \frac{1}{2}\e_{acd}{C^{cd}}_{be}\,u^e=0,\quad
E_{ab}\equiv C_{acbd} u^c u^d\neq 0,
\end{eqnarray}
has been scrutinized~\cite{Wyll}.
As a main result, it has been proved that 
the vorticity $\w^a$ of $u^a$ must be a geodesic eigenvector of
$E^a{}_b$. Moreover, the corresponding eigenvalue is linearly
related to the energy density $\mu$, which cannot be constant, as
this would lead to a set of Petrov type $D$ solutions containing the
G\"{o}del universe. Hence, if we denote $t_m$ for the number of
independent components of the Riemann tensor and its first $m$
covariant derivatives wrt the Weyl principal tetrad ($E^a{}_b$
eigenframe), we have either $t_0=1$ or $t_0=2$.
The $t_0=1$ subclass splits into two separate families: the first
consists of all non-expanding solutions ($\t=0$, for which
necessarily $\Lambda<0$), whilst the second forms a particular set,
say ${\cal S}$, of expanding models ($\t\neq 0$), all having
$\Lambda>0$.

In this communication I deduce the general line element for the
family ${\cal S}$ directly from its invariant description wrt the
Weyl principal tetrad. It depends on three free functions of one
coordinate.  The construction shows
that a metric in ${\cal S}$ admits at least three, but generically
four, functionally independent scalar invariants, and thus either
possesses a group $G_1$ of isometries or is asymmetric. In a next
step, the discussion is widened to comprise the non-rotating limit
case. Finally, denoting $q$ for the number of covariant derivatives
of the Riemann tensor required in the Karlhede invariant
classification algorithm~\cite{Karlhede2}, it is shown that
space-times belonging to a particular subfamily of ${\cal S}$ have
$q=3$. To the best of my knowledge, this is the highest value from
Petrov type $I$ examples analysed so far, the previous one being
trivial ($q=1$). In combination with the remarkable recent result by
Milson and Pelavas~\cite{Milson}, who exhibit a set of Petrov type
$N$ space-times with the theoretically maximal value $q=7$ (which
turn out to be the unique solutions with this
property~\cite{Milson2}), this reopens the question whether the
upper bound $q=5$ is sharp in the algebraically general case as
well.

\section{Line element and Karlhede classification}

For a generic member of ${\cal S}$, we write ${\cal B}\equiv
(\p_0\equiv\mathbf{ u},\p_1,\p_2,\p_3)$ for the essentially unique
Weyl principal tetrad, and  $(\W^0,\W^1,\W^2,\W^3)$ for the dual
basis of one-forms. According to the results of \cite{Wyll} we may
arrange the tetrad
such that the invariant description wrt ${\cal B}$
reduces to the following:
\begin{itemize}
\item[(i)] Curvature variables ($\l\in\mathbb{R}$):
\begin{eqnarray}
\Lambda=\lambda^2,\quad \wph=2\l(\t-\l),\label{Rab var}\\
H_{ab}=0,\quad E_{12}=E_{13}=E_{23}=0,\\
E_{11}-E_{22}=\lambda^2,\quad E_{22}-E_{33}=-\l\t,\quad
E_{33}-E_{11}=\l(\t-\l).\label{Ediff}
\end{eqnarray}
\item[(ii)] Commutator relations:
\begin{eqnarray}
[\p_1,\p_0]=[\p_1,\p_2]=0,\quad [\p_0,\p_2]=-\l\p_2,\label{com012}\\
{}[\p_0,\p_3]=-2\w\p_2-(\t-\l)\p_3,\quad
[\p_2,\p_3]=-2\w\p_0,\\
{}[\p_1,\p_3]=-\b\p_3.\label{com3}
\end{eqnarray}
\item[(iii)] Ricci and Bianchi equations:
\begin{eqnarray}
\p_0\w=-\t\w,&\quad\p_1\w=-\b\w,&\quad\p_2\w=0,\label{dw}\\
\p_0\t=-\t(\t-\l),&\quad\p_1\t=-\b(\t-\l),&\quad\p_2\t=0,\label{dtheta}\\
\p_0 \b=-(\t-\l)\b,&\quad\p_1 \b=-\b^2-\l\t,&\quad\p_2
\b=0.\label{dq}
\end{eqnarray}
\end{itemize}
Note from (\ref{Rab var}-\ref{Ediff}) that there is only one
\emph{algebraically} independent Riemann curvature component, and
that $\l\neq 0,\,\t\neq 0$ and $\t\neq \l$ by the Petrov type $I$
assumption. The describing variables $\w\equiv \w_1=\s_{23}$, $\t$
and $\b$
are invariantly defined scalars, as they
are linearly related to commutator coefficients
of the geometrically fixed tetrad ${\cal B}$. One checks that the
formal PDE system (\ref{dw}-\ref{dq})
is consistent with (\ref{com012}). 
The $\p_3$-derivatives of $\w$, $\t$ and $\b$ being undetermined,
there will be three free functions in the general solution, which we
now construct.

To start with, we note from (\ref{com012}) that $\p_0$, $\p_1$ and
$\p_2$ forms a Lie subalgebra, i.e., the vector field $\p_3$ is
normal. Hence functions $F$ and $z$ exist such that $\W^3=-\di z/F$.
$F$ is related to the commutator coefficients $\g^3{}_{jk}$ via
Cartan's first structure equation for $\W^3$,
$\di\W^3=-\frac{1}{2}\g^3{}_{jk}\W^j\wedge\W^k$,
which is equivalent to $\p_i F=\g^3{}_{i3}F$, $i<3$, i.e.,
\begin{equation}
\p_0 F=-(\t-\l)F,\quad \p_1 F=-\b F,\quad \p_2 F=0.\label{eqF}
\end{equation}
Conversely, for any solution $F$ of (\ref{eqF}) it follows that a
function $z$ exists such that $\W^3=-\di z/F$. Fixing one such
solution $F$ and defining, for any given scalar invariant $S$, the
invariant $S_F$ by
\begin{equation}
\p_3 S \equiv F S_F,
\end{equation}
the commutation relations $[\p_i,\p_3]S=\g^j{}_{i3}\,\p_j S$ read
\begin{equation}
F\p_i S_F=\p_3\p_i S+\sum_{j=0}^2\g^j{}_{i3}\,\p_j S,\quad
i<3,\label{derivSF}
\end{equation}
and the differential $\di S$ can be expanded as
\begin{equation}
\di S= \sum_{i=0}^2 \p_i S\,\W^i-S_F \di z.\label{expdS}
\end{equation}

For suitable $F$, the choice of which we postpone at the moment, the
function $z$ will serve as one of the coordinates. One observes from
(\ref{com012}-\ref{com3}) that the null vector fields
$\p_2{}\pm\p_0{}$ are also normal and that the normalized vorticity
vector $\p_1$ is normal and geodesic. However, we will not introduce
according coordinates; the deduction of the line element and the
further discussion are much more elegant and clear when we proceed
by constructing the three remaining coordinates directly from the
invariants and their $\p_3$-derivatives, hereby exploiting the
expansion (\ref{expdS}).

Firstly, we see from (\ref{dw}) and (\ref{dtheta}) that
$\p_i(\w/(\t-\l))=0$, $i<3$; it then follows from (\ref{expdS}) that
the ratio $S=\w/(\t-\l)$ is a function of $z$, which we henceforth
denote by $1/f_3(z)$ or, in view of the non-rotating limit to be
considered later, by $g_3(z)$. Next, we define the invariants $t$
and $x$ by
\begin{equation}
\tan x= -\frac{\b}{\t},\quad \cos x\, e^t = \frac{\t}{\t-\l}.
\end{equation}
This is equivalent to writing $\b$ and $\t$ in a different way,
\begin{equation}
\b= -\frac{\l\sin x\, e^t}{\cos x\, e^t-1},\quad \t = \frac{\l\cos
x\, e^t}{\cos x\, e^t-1},\label{qth}
\end{equation}
so that the vorticity amplitude and matter density read
\begin{equation}
\w=\frac{\t-\l}{f_3(z)}=\frac{\l}{f_3(z)(\cos x\, e^t-1)},\quad
\wph=2\l(\t-\l)=\frac{2\l^2}{\cos x\, e^t-1}.\label{wrho}
\end{equation}
Equations (\ref{dtheta}) and (\ref{dq}) imply the simpler
derivatives
\begin{eqnarray}
\p_0 t=\l,&\quad\p_1 t=0,&\quad\p_2 t=0,\label{dt}\\
\p_0 x=0,&\quad\p_1 x=\l,&\quad\p_2 x=0.\label{dx}
\end{eqnarray}
We now use (\ref{derivSF}). Putting $S=x$ we get $\p_i x_F=0$,
$i<3$, whence
$x_F=f_1(z)$ by (\ref{expdS}) with $S=x_F$. Putting $S=t$ and
rewriting $t_F=-2y$ we obtain
\begin{eqnarray}\label{dy}
\p_0 y=0,&\quad\p_1 y=0,&\quad\p_2 y=\frac{\l\w}{F}.
\end{eqnarray}
This is the point where we make a convenient choice for $F$. Looking
at (\ref{dw}), (\ref{eqF}) and (\ref{dt}) we see that we may pick
$F=\w e^t$, such that $\p_2 y=\l e^{-t}$ by (\ref{eqF}). Finally,
putting $S=y$ in (\ref{derivSF}) we find
\begin{eqnarray}\label{dyF}
\p_0 y_F=-2\l e^{-2t},
&\quad \p_1 y_F=0,
&\quad \p_2 y_F=2 y\l e^{-t},
\end{eqnarray}
i.e.\ $\p_i y_F=\p_i(y^2+e^{-2t})$, $i<3$, such that from
(\ref{expdS}) with $S=y_F-(y^2+e^{-2t})$ we derive
$y_F=f_2(z)+y^2+e^{-2t}$.

Assembling the above pieces one gets
\begin{eqnarray}\label{expdxi}
\begin{array}{c}
\di t =\l\W^0+2y\di z,\quad \di x =\l\W^1-f_1(z)\di z,\\
\di y =\l e^{-t}\W^2-(f_2(z)+y^2+e^{-2t})\di z, \quad
\W^3=f_3(z)(e^{-t}-\cos x)\di z.
\end{array}
\end{eqnarray}
Hence, on using $z$ and the scalar invariants $t,\,x,\,y$ as
coordinates, we infer immediately from (\ref{expdxi}) that the line
element is given by
\begin{eqnarray}\label{lineelement}
\lambda^2 \di s^2&=&-(\di t-2y\,\di z)^2+(\di x+f_1(z)\di z)^2+(\cos\,x-e^{-t})^2 f_3(z)^2 \di z^2\nonumber\\
&&+e^{2t}\left[\di y+\left(f_2(z)+y^2+e^{-2t}\right)\di z\right]^2,
\end{eqnarray}
where the invariant scalar fields $f_1(z)$, $f_2(z)$ and $f_3(z)$
are arbitrary functions of their argument. Notice that
$\lambda=\sqrt{\Lambda}$ plays the role of a constant scaling
factor. The variable $t$ is time-like and $x$ and $y$ are
space-like, whereas $z$ is spacelike, null or timelike in the region
where
\begin{equation}
g_{zz}=f_1^2+e^{2t}(f_2+e^{-2t}+y^2)^2+(\cos x-e^{-t})^2 f_3^2-4y^2
\end{equation}
is greater than, equal to or smaller than zero, respectively.
Switching to the null coordinates $u_{\pm}$ and the space-like
coordinate $\xi$ defined by
\begin{equation}\label{upm xi}
u_\pm\equiv y\pm e^{-t},\quad \xi\equiv x-\phi(z),\quad \frac{\di
\phi}{\di z}(z)\equiv -f_1(z),
\end{equation}
the metric becomes
\begin{eqnarray}
\lambda^2\di s^2=(\W^2+\W^0)(\W^2-\W^0)+(\W^1)^2+(\W^3)^2,\\
\W^2\mp\W^0=\frac{2}{u_+-u_-}[\di u_\pm+(f_2(z)+u_\pm^2)\di
z],\label{W2pmW0}\\
\W^1=\di \xi,\quad \W^3=f_3(z)\left(\cos(\xi+\phi(z))+\frac{u_-
-u_+}{2}\right)\di z.
\end{eqnarray}
(\ref{W2pmW0}) makes the normality of $\W^2\mp\W^0$ apparent, as
$\di u_\pm+(f_2(z)+u_\pm^2)\di z$ are one-forms on 2-spaces; at the
same time it reveals the difficulty when we would have started with
coordinates $v_\pm$ and scalar fields $G_\pm(v_+,v_-,\xi,z)$ for
which $\W^2\mp\W^0=G_\pm \di v_\pm$.

One observes from (\ref{qth}) and (\ref{wrho}) that $\wph>0$, and at
the same time $\t>0$, at space-time points with $\cos x>e^{-t}$,
while the boundary $\cos x=e^{-t}$ forms a two-surface of curvature
singularities. It is an open question whether (\ref{lineelement})
may generate physically plausible rotating dust models.

The discussion of possible Killing vector fields becomes 
trivial in the above invariant approach: when at least one of the
functions $f_1(z), f_2(z)$ or $f_3(z)$ is non-constant it can be
taken as a fourth invariantly defined coordinate replacing $z$, and
there is no symmetry; when all of them are constant there is a group
$G_1$ of
isometries generated by $\partial/\partial z$.\\

By putting $\w$ equal to zero in (\ref{Rab var}-\ref{dq}) one
arrives at the corresponding irrotational dust family ${\cal S}^0$,
which was integrated in \cite{VdBWyll1} (metric (45)). More
generally, we may consider the class $\widetilde{{\cal S}}={\cal S}
\cup {\cal S}^0$.
Then the above integration procedure can be applied  up to
(\ref{dy}), and we make the alternative choice $F=(\theta-\l)\,e^t$.
Taking further advantage of the normality of e.g.\ $\p_2-\p_0$ and
solving Cartan's first structure equations, one readily obtains the
line element
\begin{eqnarray}\label{lijn}
\l^2 \di s^2=&&-(\di t-2y\,\di z)^2+(\di x+f_1(z)\di z)^2+(\cos
x-e^{-t})^2\di z^2\nonumber\\&&+(e^{t-K}\di \eta+\di t-2y\,\di z)^2,
\end{eqnarray}
where now
\begin{equation}\label{y}
y=y(t,\eta,z)=\frac{1}{2}\fpd{K}{z}(\eta,z)-e^{-t}g_3(z)
\end{equation}
and where $K=K(\eta,z)$ is a solution of
\begin{equation}\label{bond}
e^{K(\eta,z)}\frac{\p^2 K}{\p \eta\, \p z}(\eta,z)=2g_3(z).
\end{equation}

The non-rotating subcase corresponds to $g_3(z)=0$. Looking at
(\ref{dy}), (\ref{y}) and (\ref{bond}), and after possibly
redefining $\eta$, we have that $K=K(z)$ is a primitive function of
$2y=2y(z)$.
In terms of coordinates $z$, $\tau\equiv t-K(z)$, $\zeta \equiv
\eta-e^{-t}$ and $\xi$, cf.\ (\ref{upm xi}), the non-rotating
solutions take the form
\begin{eqnarray}\label{line nonrot}
\lambda^2 \di s^2&=&-\di \tau^2+\di\xi^2+e^{2\tau}\di
\zeta^2+[e^{-K(z)}e^{-\tau}-\cos(\xi+\phi(z))]^2\di z^2.
\end{eqnarray}
There is at least a $G_1$ isometry group generated by $\p_\zeta\sim
e^\tau\p_2$. As the scalar invariants $t$, $x$, $y$ and $f_1$ must
be invariant under any isometry, one immediately deduces that there
is a group $G_2$ of motions if and only if both $y$ and $f_1$ are
constant, i.e., if and only if the line element can be transformed
to (\ref{line nonrot}) with $K(z)=yz$ and $\phi(z)=f_1 z$, where $y$
and $f_1$ are constants; the additional Killing vector field is then
given by
\begin{equation}
2y\p_\tau-f_1\p_\xi-\p_z\sim 2y\p_0-f_1\p_1+(e^{-t}-\cos x)\p_3.
\end{equation}
This corrects an inaccuracy in the symmetry discussion of
${\cal S}^0$ in \cite{VdBWyll1}.\\

When working wrt the Weyl principal tetrad ${\cal B}$ of a Petrov
type $I$ space-time, the stop value $q$ within the Karlhede
invariant classification algorithm is defined by 
\begin{equation}
t_m>t_{m-1},\,0\leq m<q, \quad t_q=t_{q-1},
\end{equation}
where one formally puts $t_{-1}\equiv -1$. The final value $t_{q-1}$
is the total number of functionally independent scalar invariants
and equals $4-r$, where $r$ stands for the dimension of the isometry
group. For members of $\widetilde{{\cal S}}$ one has $t_0=1$ and
$t_{q-1}=2,3$ or $4$, such that $2\leq q\leq 4$ a priori. A direct
calculation shows that the components of the first covariant
derivative of the Riemann tensor wrt ${\cal B}$ are rational
functions of the invariants
\begin{equation}
\cos x \,e^t,\quad \sin x \,e^t,\quad 2\,y\cos \,x +\sin
x\,f_1(z),\quad g_3(z).
\end{equation}
The following possibilities thus arise
($f'=0$ means that $f$ is constant):
\begin{itemize}
\item for metrics in ${\cal S}^0$ ($g_3(z)=0$, $y=y(z)$):
\begin{enumerate}
\item $f_1'=y'=0$. There is a group $G_2$ of
isometries and $t_1=2$, such that $t_2=2$ and $q=2$.
\item $f_1'\neq 0$ or $y'\neq 0$. There is a group $G_1$ of
isometries and $t_1=3$, such that $t_2=3$ and $q=2$.
\end{enumerate}
\item for metrics in ${\cal S}$ ($g_3(z)=1/f_3(z)\neq 0$, $y$ functionally independent of $t$, $x$ and $z$):
\begin{enumerate}
\item $f_1'=f_2'=f_3'=0$. There is a
group $G_1$ of isometries and $t_1=3$, such that $t_2=3$ and $q=2$.
\item $f_3'=0$, but $f_1'\neq 0$ or $f_2'\neq 0$. There
is no symmetry and $t_1=3$, such that $t_2=t_3=4$ and $q=3$.
\item $f_3'\neq 0$. There is no symmetry and $t_1=4$, such that $t_2=4$ and $q=2$.
\end{enumerate}
\end{itemize}
We conclude that $q=2$ for all members of $\widetilde{{\cal S}}$,
except for the asymmetric (whence rotating) models for which the
ratio $2\l f_3=\wph/\w$ of energy density and vorticity amplitude is
constant. This special class of models, depending on two free
functions of one coordinate and one constant parameter, provides a
first example of algebraically general space-times for which $q=3$.

\ack I would like to thank Norbert Van den Bergh for suggesting this
integration technique, discussions on the subject and careful
reading of the document, and Jan
Aman for verifications in the CAS package CLASSI. The GRTensorII
package has been used to a posteriori check the invariant properties
of the line elements.


\section*{References}

\begin{thebibliography}{40}

\bibitem{Winicour} Winicour J \Journal{J. Math. Phys.}{16}{1806}{1975}
\bibitem{SKMHH} Stephani H, Kramer D, MacCallum M A H, Hoenselaers C and Herlt E 2003,
\emph{Exact Solutions to Einstein's Field Equations, Second Edition}
(Cambridge: Cambridge University Press)
\bibitem{Stephani} Stephani H \Journal{Class. Quantum Grav.}{4}{125}{1987}
\bibitem{Barnes} Barnes A \Journal{Class. Quantum Grav.}{16}{919}{1999}
\bibitem{Godel} G\"{o}del K \Journal{Rev. Mod. Phys.}{21}{447}{1949}
\bibitem{Wyll} Wylleman L, \emph{Algebraically general,
gravito-electric rotating dust}, \emph{Preprint} gr-qc 0804.3222
\bibitem{Karlhede2} Karlhede A \Journal{Gen. Rel. Grav.}{12}{693}{1980}
\bibitem{Milson} Milson R and Pelavas N \Journal{Class. Quantum Grav.}{25}{012001}{2008}
\bibitem{Milson2} Milson R and Pelavas N, \emph{Preprint} gr-qc 0711.3851
\bibitem{VdBWyll1} Van den Bergh N and Wylleman L \Journal{Class. Quantum Grav.}{21}{2291}{2004}

\end{thebibliography}
\end{document}